\documentclass[11pt,preprint,preprintnumbers,amsmath,amssymb,nofootinbib]{revtex4}
\usepackage{color}
\usepackage{graphicx,graphics}

\newcommand{\be}{\begin{equation}}  
\newcommand{\ee}{\end{equation}}  
\newcommand{\bea}{\begin{eqnarray}}  
\newcommand{\eea}{\end{eqnarray}}  
\newcommand{\ba}{\begin{array}}  
\newcommand{\ea}{\end{array}}

\begin{document}
\preprint{FERMILAB-PUB-20-499-T}
\preprint{OUTP-20-10P}

\title {Gravitational Contact Interactions 
and the Physical Equivalence \\ 
of Weyl Transformations {in Effective Field Theory}}


\author{Christopher T. Hill}
\email{hill@fnal.gov}
\affiliation{Fermi National Accelerator Laboratory\\
P.O. Box 500, Batavia, Illinois 60510, USA\\$ $}
\author{Graham G. Ross}
\email{g.ross1@physics.ox.ac.uk}
\affiliation{Rudolf Peierls Centre for Theoretical Physics, \\
University of Oxford, 1 Keble Road\\
Oxford OX1 3NP\\$ $}

\date{\today}

\begin{abstract}
Theories of scalars and gravity, with non-minimal interactions,
$\sim (M_P^2 + F(\phi_i) )R +L(\phi_i)$,
have graviton exchange induced contact terms.
These terms arise in single particle reducible diagrams with vertices, $\propto q^2$, that cancel
the Feynman propagator denominator, $1/q^2$, and are familiar in various other physical contexts. 
In gravity these lead
to additional terms in the action such as $\sim F(\phi_i) T_\mu^\mu(\phi_i)/M_P^2$
and $F(\phi_i) \partial^2 F(\phi_i)/M_P^2$.
The contact terms are equivalent to induced operators obtained by a Weyl transformation
that removes the non-minimal interactions, leaving a minimal Einstein-Hilbert gravitational action.
This demonstrates explicitly the equivalence of different representations of the action
under Weyl transformations, both classically and quantum mechanically. To avoid {such} ``hidden
contact terms'' one is compelled to go to the minimal Einstein-Hilbert representation.
\end{abstract}

\maketitle

\maketitle
\section{Introduction}

In recent years there has been considerable interest in scale invariant theories that,
by way of spontaneous scale symmetry breaking or ``inertial symmetry breaking,''
dynamically generate the Planck mass and associated phenomena
of inflation and hierarchies \cite{Wetterich1,authors,inertial}.  A feature many of these approaches
 have in common is the notion of some pre-Planckian era, in which fundamental scalars
 exist and couple to
gravity through non-minimal interactions, $\sim F(\phi_i)R$.  The scalars then acquire
VEV's that lead to a Planck mass, $\sim M_P^2 R + F'(\phi_i)R$, where $F'$
contains residual active scalar fields that couple non-minimally. 

A key tool in the analysis of these models is the Weyl transformation, \cite{Weyl}.  This involves
a redefinition of the metric, $ g' =\Omega(\phi_i)g$, in which $g$ comingles 
with scalars. $\Omega$ can be chosen to 
lead to a new effective theory, typically one that is pure Einstein-Hilbert,
  $\sim M_P^2 R + L'(\phi_i)$, in which the non-mimimal interactions 
  have been removed.\footnote{Alternatively, one might partially remove a subset of scalars from the non-minimal interactions $\sim M_P^2 R + F''(\phi_i)R$ where $F''$ is optimized for some particular model
application.}

The Weyl transformation is classically exact. However, it is often 
difficult to
discern how the original non-minimal interaction theory is physically equivalent to the
pure Einstein-Hilbert form. There may be apparent
advantages in using the transformed theory that are not evident in the original, or vice versa.
These apparent
advantages, however, may not really be present 
when all effects are taken into account.
It is also unclear how the Weyl transformation is compatible with a full quantum theory \cite{Duff}.

In the present paper we will address these questions. We will work to first order in $1/M_P^2$
in a linearized version of a theory with Planck mass and non-minimal interactions. We will
not perform a Weyl redefinition of the metric.  Nonetheless, we will demonstrate how the 
Weyl transformation form of $L'(\phi_i)$ necessarily arises perturbatively by way of Feynman diagrams involving graviton exchange.

This happens by way of {\em contact terms} that are generated by the graviton exchange
amplitudes.  These are bona fide physical effects that occur in various venues
in physics and, though they arise in tree approximation,
they must be included into
the effective action of the theory at the given order of perturbation theory. 
Moreover, this represent essentially ``integrating
out'' the vertices that lead to the contact terms.  The result is that the non-minimal
interactions will disappear from the theory at any given order in perturbation
theory and are replaced by new, pointlike 
interactions from the contact terms.  

Perhaps not surprisingly, the form of the contact
term interactions corresponds identically with
the Weyl transformation that takes the theory to the pure Einstein-Hilbert form.
We argue that,  once the Planck scale is generated in the theory,
by spontaneous or ``inertial''
 symmetry breaking \cite{inertial}, then the action should be ``diagonalized,'' 
 in analogy to diagonalizing the kinetic terms, so that
 the contact terms do not appear perturbatively.  This mandates a Weyl transformation to a
 pure Einstein-Hilbert action which is a unique specification of the theory. 
 
 We will be computing potentials that arise from graviton exchange. This will
 require gauge fixing, and we will use the standard De Donder gauge in a first pass,
 following Donoghue, et.al. \cite{Donoghue}. However,
 we will also find it illuminating to consider a different gauge choice which
 separates a traceless metric from it's trace. The trace metric has
 a ghost signature, but it uniquely controls the relevant contact terms associated
 with the Weyl transformation. Otherwise, both gauges give the same results, as they must.
 
 We turn presently to a brief discussion of contact terms in general and
 a toy model
 that will be structurally similar to the gravitational case.

\subsection{Contact Interactions}

Generally, single particle irreducible (1PI) Feynman diagrams describe 
perturbative  corrections (or renormalizations)
of a Lagrangian based field theory action. On the other hand, reducible diagrams, those that break into two disconnected
diagrams upon cutting a line,
are the radiative effects that one computes from the given action \cite{Jackiw}.  There is, however, an exception: sometimes single particle reducible diagrams correspond to ``contact term'' interactions. These then become part of the action.

Contact terms arise in a number of
phenomena.   Diagrammatically they arise when a
vertex for the emission of, e.g., a massless  quantum, of momentum $q_\mu$,
is proportional to $q^2$.  This vertex then cancels the $1/q^2$
from the propagator when the quantum is exchanged. This $q^2/q^2$
cancellation leads to an
effective pointlike operator from an otherwise single-particle reducible diagram.

For example, in electroweak physics a
vertex correction by a $W$-boson to a gluon emission 
induces a quark flavor changing operator, e.g.,
describing $s\rightarrow d+$gluon, where $s$ ($d$) is a strange (down) quark.
This has the form of a local operator:
\bea
\label{one}
g\kappa \bar{s}\gamma_\mu T^A d_L D_\nu G^{A\mu\nu}
\eea
where $G^{A\mu\nu}$
is the color octet
gluon field strength and $\kappa \propto
G_{Fermi}$.
This implies a vertex 
for an emitted gluon of 4-momentum $q$ and polarization $\epsilon^{A\mu}$,
of the form
$g\kappa\bar{s}\gamma_\mu T^A d_L  \epsilon^{A\mu}\times q^2 +...$.  However,
the gluon propagates, $\sim 1/q^2$, and couples to a quark current 
$\sim g\epsilon^{A\mu}\bar{q}\gamma_\mu T^A q$.
This results in a contact term:
\bea
\label{localop}
g^2 \kappa\left(\frac{q^2}{q^2}\right) \bar{s}\gamma^\mu T^A d_L \bar{q}\gamma_\mu T^A q
\;\sim\;
g^2 \kappa\bar{s}\gamma^\mu T^A d_L\bar{q}\gamma_\mu T^A q
\eea
The result is a 4-body local operator 
which mediates electroweak transitions
between, e.g., kaons and pions \cite{Shifman}, also
known as ``penguin diagrams'' \cite{penguins}.
Note the we can rigorously obtain the contact term result
by use of the gluon field equation within the operator
of eq.(\ref{one}), 
\bea
D_\nu G^{A\mu\nu}= g\bar{q}\gamma^\mu T^A q.
\eea
This is justified as operators that vanish by equations of motion, 
known as ``null operators,'' will generally have
gauge noninvariant anomalous dimensions and are unphysical 
\cite{Deans}.

Another example of a contact term occurs in the case of
a cosmic axion, described by an oscillating  classical 
field,  $\theta(t)=\theta_0\cos(m_at)$, 
interacting with a magnetic moment, $ \vec{\mu}(x)\cdot \vec{B}$, through the
electromagnetic anomaly $ \kappa \theta(t) \vec{E}\cdot \vec{B} $. A static magnetic moment 
emits a virtual spacelike photon of momentum $(0,\vec{q})$. The anomaly 
absorbs the  virtual photon and emits an on-shell photon of polarization $\vec{\epsilon}$,
inheriting energy $\sim m_a$ from the cosmic axion.
The Feynman diagram, with the exchanged virtual
photon, yields an amplitude, 
$\propto (\theta_0\mu^i\epsilon_{ijk}q^j)(1/\vec{q}^{\;2}) (\kappa \epsilon^{k\ell h}q_\ell m_a\epsilon_h)
\sim  (\kappa\theta_0 m_a\vec{q}^{\;2}/\vec{q}^{\;2}) \vec{\mu}\cdot \vec{\epsilon}$. 
The $ \vec{q}^{\;2}$ factor then 
cancels the $1/\vec{q}^{\;2}$ in the photon propagator,
resulting in a contact term which is an induced, parity violating,
oscillating electric dipole interaction:
$\sim \kappa \theta (t) \vec{\mu}\cdot \vec{E}
$. This results in 
cosmic axion induced {\em electric dipole} radiation from
any magnet, including an electron \cite{CTHa}.

\subsection{Illustrative Toy Model of Contact Terms}

In preparation for the analysis of gravitational contact terms we first present a schematic discussion of a simple toy model that illustrates the emergence of contact terms and is structurally similar
to what we encounter in gravity.\footnote{Here Lorentz indices have been suppressed and the contraction of indices understood.
$\widehat{T}$ refers to the time ordered product, where $T$ is the trace of the stress tensor.} Consider a single real scalar field $\phi $
and operators $A$ and $B,$ which can be functions of other fields, with the action given by:
\bea
S=\int \frac{1}{2}\partial \phi \partial \phi -A\partial ^{2}\phi -B\phi 
\label{toy}
\eea
Here
$\phi $ has a propagator $\frac{i}{q^{2}}$, but the vertex of a diagram
involving
$A$ has a factor of $\partial^2 \sim -q^{2}$. This yields a pointlike interaction, $\sim
q^{2}\times \frac{i}{q^{2}}$,
in a single particle exchange of $\phi $, and therefore implies contact terms: 
\bea
 \!\! \widehat{T} \;\;  i\!\!\int \! A\partial ^{2}\phi \;\;\; i\!\!\int \! B\phi & \rightarrow & -\frac{i}{q^{2}}\left(
-q^{2}\right) AB\;\;\;=\;\;\; i\!\!\int \! AB\nonumber \\
\frac{1}{2}\widehat{T} \;\;  i\!\!\int \! A\partial ^{2}\phi \;\;\;
i\!\!\int \! A\partial ^{2}\phi \;\;\;
 & \rightarrow & -
\frac{i}{2q^{2}}A^2 \left( -q^{2}\right)^2  \;\;=\;\;\frac{i}{2}
\int \! A\partial ^{2}A.
\eea
This also produces  a nonlocal  interaction $ -\frac{i}{2q^{2}}BB.$ 

\begin{figure}[t!]
\vspace{0.0 in}
	\hspace*{1.0in}\includegraphics[width=\textwidth]{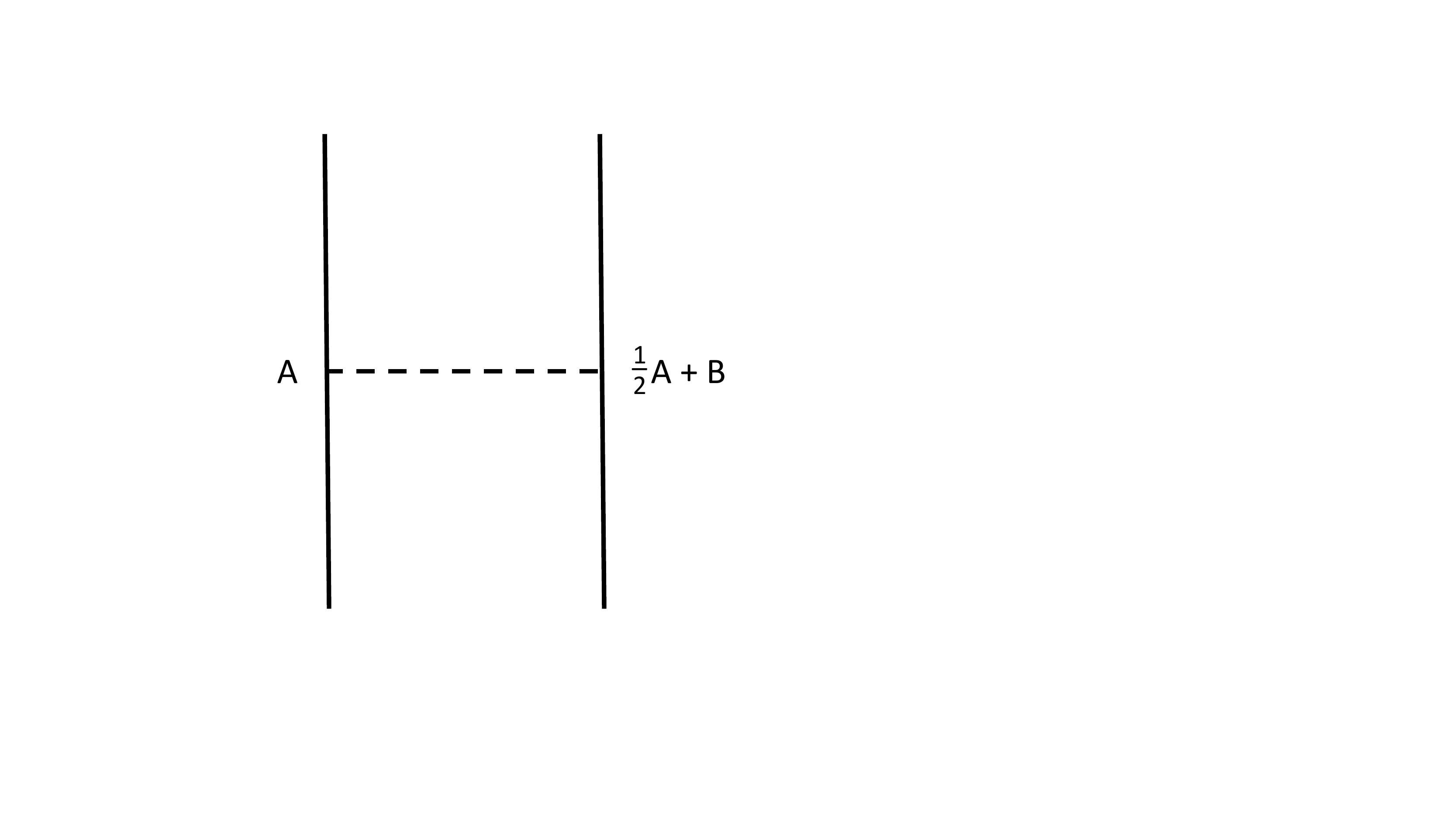}
	\vspace{-1.0 in}
	\caption{Contact terms in the toy model are
	generated by diagrams with exchange of $\phi$ (dashed). In gravity,
	with non-minimal term $\sim \int \sqrt{-g} F\left( \phi _{i}\right) R$ 
	and matter field Lagrangian $\sim \int \sqrt{-g} L\left( \phi _{i}\right)$
	then $A$ is replaced by $F(\phi)$ and $B$ is replaced by $L(\phi)$, and the dashed line
	is a graviton propagator. }
\end{figure}

Exponentiating these operators we see that we have
diagrammatically obtained a local effective action:
\bea
S=\int \frac{1}{2}\partial \phi \partial \phi +\frac{1}{2}A\partial
^{2}A+AB +\makebox{ {long distance terms}}
\eea
Of course, we can see this straightforwardly by
``solving the theory,'' by defining a shifted field:
\bea
\phi =\phi^{\prime } -\frac{1}{\partial ^{2}}\left( \partial ^{2}A+B\right) 
\eea
Substituting and integrating by parts, this yields:
\bea
\label{eff1}
S =\int \frac{1}{2}\partial \phi ^{\prime }\partial \phi ^{\prime }+\frac{%
1}{2}A\partial ^{2}A+AB+\frac{1}{2}B\frac{1}{\partial ^{2}}B
\eea
An equivalent effective local action that describes both
short and large distance is then,
\bea
\label{eff2}
S=\int \frac{1}{2}\partial \phi \partial \phi +\frac{1}{2}%
A\partial ^{2}A+AB-{B \phi }
\eea
The contact terms have become pointlike components of the effective action,
while the
{long} distance effects are produced by $\phi $ exchange.  Note that the
derivatively coupled
operator $A$ has no long distance interactions {due to} $\phi $ exchange. 
Moreover, in the
effective action of
eq.(\ref{eff2}) we have implicitly ``integrated out'' the $A\partial ^{2}\phi $, which is no
longer part of the action and is replaced
by new operators $\frac{1}{2}A\partial ^{2}A+AB$.
One can also adapt the use of equations of motion to simplify the action but
this requires care. For example, the insertion of the $\phi$ equation of motion
into $A\partial^2\phi$ correctly gives the $AB$ term but misses the 
factor of $1/2$ in the $A\partial^2A$ term.

\section{Gravitational Contact Terms}

We will consider a general theory involving scalar fields $\phi _{i},$
an Einstein-Hilbert {term and} a non-minimal interaction:
\bea
S=\int \sqrt{-g}\left(\frac{1}{2}M_{P}^{2}R\left( g_{\mu \nu }\right) +\frac{1}{2%
}F\left( \phi _{i}\right) R\left( g_{\mu \nu }\right) +L\left( \phi
_{i}\right) \right).
\eea
where we use the 
metric signature and curvature tensor conventions of \cite{CCJ}.
In parallel with the general discussion we will quote
the results {for} a simple model,
\bea
S=\int \sqrt{-g}\left(\frac{1}{2}M_{P}^{2}R\left( g_{\mu \nu }\right) +\frac{1}{2%
}\xi \phi ^{2}R\left( g_{\mu \nu }\right) +\frac{1}{2}g^{\mu \nu }\partial
_{\mu }\phi \partial _{\nu }\phi -W(\phi )\right).
\eea
The matter lagrangian has {the} stress tensor and stress tensor trace:
\bea
T_{\mu \nu }&=& \frac{2}{\sqrt{-g}}\frac{\delta }{\delta g^{\mu \nu }}%
\int \sqrt{-g}L\left( \phi _{i}\right)
\nonumber \\
T &=& g^{\mu\nu}T_{\mu\nu}
\eea
which, in the simple model, take the form,
\bea
T_{\mu \nu }&=&\partial _{\mu }\phi \partial _{\nu }\phi -g_{\mu \nu }\left( 
\frac{1}{2}g^{\rho \sigma }\partial _{\rho }\phi \partial _{\sigma }\phi
-W(\phi )\right) .
\nonumber \\
T&=&-\partial ^{\sigma }\phi \partial _{\sigma
}\phi +4W(\phi )
\eea
This is the usual matter stress tensor and it is conserved
by the $\phi $ equations of motion to leading order in $1/M_P^2$
in  a linearized gravity approximation, and we can neglect
the contribution of the non-minimal
term ($S_2$ below) to the stress tensor conservation at this order.\footnote{
This is not the ``improved stress tensor'' of 
\cite{CCJ},
where ``improvement terms'' are separately explicitly conserved
and come from an assumed conformal non-minimal coupling of $\phi $ to gravity,$\ 
\frac{1}{2}\xi\phi ^{2}R,$ with
 $\xi =\frac{1}{6},$ and is not relevant in the present
discussion.}

We treat the theory perturbatively, expanding around flat space. {Hence} we
linearize gravity with a weak field $h_{\mu \upsilon }$:
{\bea
\label{linear}
g_{\mu \upsilon }\approx \eta _{\mu \upsilon }+\frac{h_{\mu \upsilon }}{%
M_{P}},\qquad g^{\mu \upsilon }\approx \eta ^{\mu \upsilon }-\frac{h^{\mu
\upsilon }}{M_{P}}+O(h^{2}),\qquad \sqrt{-g}\approx 1+\frac{1}{2}\frac{h}{%
M_{P}},\qquad h=\eta^{\mu\nu}h_{\mu\nu}
\eea}
The scalar curvature is then:
\bea
R &=& R_{1}+R_{2}
\nonumber \\
M_P R_{1}&=&\left( \partial ^{2}h\ -\partial^{\mu }\partial ^{\nu }h_{\mu \nu }\right) 
\nonumber \\
M_{P}^{2}R_{2}&=&  -\frac{3}{4}\partial ^{\rho }h^{\mu
\nu }\partial_{\rho}h_{\mu \nu }-\frac{1}{2}h^{\mu
\nu }\partial ^{2}h_{\mu \nu }-\frac{1}{2}h^{\mu \nu
}\partial _{\mu }\partial _{\nu }h
\nonumber \\
&&
+\partial _{\nu }\left( h^{\nu \mu }\partial ^{\rho
}h_{\mu \rho }\right) -\frac{1}{2}\partial _{\nu }\left(
h^{\nu \mu }\partial _{\mu }h\right) +h^{\mu \nu }\partial
_{\rho }\left( \partial _{\mu }h_{\nu }^{\rho }\ -\frac{1}{2}%
\partial ^{\rho }h_{\mu \nu}\right) 
\nonumber \\
&&
 +\frac{1}{2}\partial _{\mu }h^{\mu \rho }\partial _{\nu
}h_{\rho }^{\nu }-\frac{1}{2}\partial _{\mu }h\partial ^{\nu }h_{\nu
}^{\mu }\ +\frac{1}{4}\partial _{\mu }h\partial ^{\mu }h  
\eea
{Using this the action is given by 
$ S= S_{1}+S_{2}+S_{3} $}
where  $S_{1}$ is the Fierz-Pauli action:
\bea
S_{1} &=& \frac{1}{2}M_{P}^{2}\int \sqrt{-g}R
= \frac{1}{2}%
M_{P}^{2}\int \left( R_{2}+\frac{1}{2}\frac{h}{M_{P}}R_{1}\right) 
\nonumber \\
&=& \frac{1}{2}\int h^{\mu \nu }\left( \frac{1}{4}\partial ^{2}\eta _{\mu \nu
}\eta _{\rho \sigma }-\frac{1}{4}\partial ^{2}\eta _{\mu \rho }\eta _{\nu
\sigma }-\frac{1}{2}\partial _{\rho }\partial _{\sigma }\eta _{\mu \nu }+%
\frac{1}{2}\partial _{\mu }\partial _{\rho }\eta _{\nu \sigma }\right)
h^{\rho \sigma }
\eea
 Note that the leading term in the first order expansion $\frac{1}{2}M_{P}^{2}R_{1}$
is a total divergence and is zero in the Einstein-Hilbert
action. What remains
is the Fierz-Pauli action written in a factorized form
$h\left( ...\right) h$.

On the other hand the non-minimal interaction, $S_{2}$, takes the form:
\bea
S_{2}=\frac{1}{2}\int \sqrt{-g}F\left( \phi _{i}\right) R\left( g_{\mu \nu }\right) =
\int \frac{1}{2M_{P}}F\left( \phi _{i}\right) \; \Pi
^{\mu \nu }h_{\mu \nu } 
\eea
where it is useful to introduce the transverse derivative,
\bea
\Pi ^{\mu \nu }= \partial ^{2}\eta ^{\mu \nu }\ -\partial
^{\mu }\partial ^{\nu } .
\eea
Finally, $S_{3}$ is the matter action and coupling to the gravitational weak
field:
\bea
S_{3}=\int \;L\left( \phi _{i}\right) -\frac{h^{\mu \nu }}{2M_P}T_{\mu \nu }
\eea
Due to the conservation of $T^{\mu \nu }$ and the transverse derivative,
the full action $S$ possesses the local gauge invariance,
\bea
\delta h_{\mu
\nu }=\partial _{\mu }A_{\nu }+\partial _{\nu }A_{\mu }.
\eea

{Since $S_{2}$ involves derivatives, 
the} Feynman diagrams involving $S_{2}$ and $S_{3}$ will generate
contact terms in the gravitational potential generated by single graviton
exchange. This will closely parallel the toy model.

\subsection{ Graviton Propagator}

We are interested in the gravitational potential
amongst the operators that comprise $S_{2}$ and $S_{3}$. This is mediated
by a
single graviton exchange, as in Figure 1,
 effectively integrating out the
$S_{2}$ term in analogy to the $A\partial ^{2}\phi $ term in the toy model.
For this we require the graviton propagator and, due to the underlying gauge invariance, it is necessary first to gauge-fix. 

A conventional choice of gauge is the
De Donder gauge:\footnote{We presently follow the lecture notes of
Donoghue {\it et. al.,} \cite{Donoghue}, though we differ in normalization;
note the correspondence of our normalization
to Donoghue's \cite{Donoghue} is $\kappa=2/M_P$.}
\bea
\label{de}
\partial _{\mu }h^{\mu \nu }=\frac{1}{2}\partial
^{\nu }h
\eea
which is defined by the condition,
\bea
0=g^{\mu \nu }\Gamma _{\mu \nu }^{\lambda }=\frac{1}{2}\left(
\eta ^{\mu \nu }\partial _{\mu }h_{\nu }^{\lambda }\ +\eta
^{\mu \nu }\partial _{\nu }h_{\mu }^{\lambda }-\partial ^{\lambda
}\eta ^{\mu \nu }h_{\mu \nu }\right) .
\eea
The De Donder gauge is a member of a one-parameter family of gauges 
defined by $\partial _{\mu }h^{\mu \nu }=w\partial
^{\nu }h$, where $w=\frac{1}{2}$ in the De Donder case.
In Section IV we  discuss an alternative
gauge, $w=\frac{1}{4}$, which is somewhat more
transparent for our present application but,
of course, yields the same results.

The Fierz-Pauli action in De Donder gauge, by substituting eq.(\ref{de})
into $S_{1}$, takes the form:
\bea
S_{1}=\frac{1}{2}\int \frac{1}{8}h^{\mu \nu }\left( \eta _{\mu \nu }\eta
_{\rho \sigma }-\eta _{\mu \rho }\eta _{\nu \sigma }-\eta _{\nu \rho }\eta
_{\mu \sigma }\right) \partial ^{2}h^{\rho \sigma }
=\frac{1}{2}\int h^{\mu \nu }\left(\frac{1}{8} P_{\mu \nu \text{ }\rho \sigma }\right)
\partial ^{2}h^{\rho \sigma }
\eea
where,
\bea
P_{\mu \nu \text{ }\rho \sigma }=\eta _{\mu \nu }\eta _{\rho \sigma
}-\eta _{\mu \rho }\eta _{\nu \sigma }-\eta _{\nu \rho }\eta _{\alpha \sigma }
\eea
and 
$P_{\mu \nu  \text{ }\rho \sigma }$ is the spin-2 projection operator.

The inverse of the kinetic term operator is  $A^{\mu \nu \text{ }\rho \sigma } $, given by: 
\bea
&&
\frac{1}{8}\left( \eta _{\mu \nu }\eta _{\alpha \beta }-\eta _{\mu \alpha
}\eta _{\nu \beta }-\eta _{\mu \beta }\eta _{\nu \alpha }\right) A^{\alpha
\beta \text{ }\rho \sigma }=\delta _{\mu \nu }^{\rho \sigma }
\nonumber \\
&&
\delta _{\mu \nu }^{\rho \sigma }=\frac{1}{2}\left( \delta _{\mu }^{\rho
}\delta _{\nu }^{\sigma }+\delta _{\nu }^{\rho }\delta _{\mu }^{\sigma
}\right) 
\nonumber \\
&&\makebox{hence,  }\;\;
A^{\mu \nu \text{ }\rho \sigma }=2P^{\mu \nu \text{ }\rho \sigma }.
\eea
Note that the normalization follows from our
choice of scale, $\sim h_{\mu\nu}/M_P$, in the linear gravity expansion, eq.(\ref{linear}). 
This gives the propagator in a path integral with action $S_{1}$:
\be
\label{gprop}
\langle 0|\widehat{T}\text{ }h_{\rho \sigma }(x)\text{ }h_{\mu \nu
}(y)\;|0\rangle =\int Dg\; e^{iS_{1}}\left( h_{\rho \sigma }(x)h_{\mu
\nu }(y)\right) =iA_{\mu\nu\text{ }\rho\sigma}D(x-y)
\ee
where 
\bea
D(x-y)=\frac{1}{\partial ^{2}}=\int \frac{-1}{q^{2}+i\epsilon }%
e^{iq\cdot (x-y)}\frac{d^{4}q}{\left( 2\pi \right) ^{4}}
\eea
is a time ordered scalar field Green's function
satisfying $\partial ^{2}D(x-y)=\delta ^{4}(x-y).$\footnote{We use the
shorthand $\frac{1}{\partial^2}f(x) = \int D(x-y)f(y)d^4y$, and
$\frac{1}{\partial^2\partial^2}f(x) = \int D(x-y)D(y-z)f(z)d^4yd^4z$, etc.}
The
momentum space Feynman propagator for gravitons is then,
\bea
\langle 0|\widehat{T}\text{ }h_{\rho \sigma }(q)\text{ }h_{\alpha \beta
}(-q)\;|0\rangle  &=& \frac{-i}{q^{2}+i\epsilon }A_{\rho \sigma ,\alpha
\beta } .
\label{gprop2}
\eea

The procedure of substituting the gauge condition into the
action, then inverting, is analogous in electrodynamics to substituting $%
\partial _{\mu }A^{\mu }=0$
into the action, which yields the photon propagator in Feynman gauge, $\sim
-ig_{\mu \nu }/q^{2}.$
In analogy
to using Feynman gauge, we must take care to tie the graviton propagator, eq(\ref{gprop}), onto
conserved currents, such as the stress tensor or the transverse derivative,
which then guarantees gauge invariance of a given tree amplitude.

\subsection{Newtonian Potential}

Let us first consider the Newtonian potential.  This can
be computed from a Feynman diagram for graviton exchange.
Equivalently, the action is determined by simply shifting the  graviton
field.
Using the truncated action:
\bea
S= \int \frac{1}{2}h^{\mu \nu }\left( \frac{1}{8}
P_{\mu \nu \text{ }\rho \sigma }\right) \partial ^{2}h^{\rho \sigma }-%
\frac{h^{\mu \nu }}{2M_P}T_{\mu \nu }
\eea
we can define a shifted $h^{\prime \rho \sigma }$:
\bea
h^{\rho \sigma }=
h^{\prime \rho \sigma }+\frac{1}{2M_P}%
\frac{1}{\partial ^{2}}A^{\mu \nu \text{ }\rho \sigma }T_{\mu \nu }
\eea
Hence,
\bea
\label{pot1}
S&=&\frac{1}{2}\int h^{^{\prime }\mu \nu }\left(\frac{1}{8}
P_{\mu \nu \text{ }\rho \sigma }\right) \partial ^{2}h^{\prime \rho
\sigma }
-\frac{1}{2}\left( \frac{1}{2M_P}\right) ^{2}\int \int d^{4}x\; d^{4}y\;
T^{\mu \nu }(x)A_{\mu \nu \rho \sigma }D(x-y)T^{\rho \sigma}(y).
\nonumber \\
\eea
For stationary masses, located at $x=0$ and $x=r$
the stress tensor is pure $00$,
\bea
T^{00 }(x)=m_1\delta^3(\vec{x})+m_2\delta^3(\vec{x}-\vec{r}).
\eea
Insert this into the second term of eq.(\ref{pot1}), and note 
the time integrated stationary Green's function becomes,
\bea
\int \int d^{4}x\; d^{4}y\; \delta^3(\vec{x}) \;\delta^3(\vec{y}-\vec{r})\; D(x-y)=\int dt\; \frac{1}{4\pi r}
\eea
and $A_{0000}=-2$, which yields the effective action,
\bea
\label{newt}
\int \frac{2}{(2M)^2}\frac{1}{4\pi r}m_{1}m_{2}\;dt =\int \frac{G_N m_1 m_2}{r} dt
\eea
where $ M_P^{2}=(8\pi G_{N})^{-1}$
and implies an attractive Newtonian gravitational potential.

The Feynman propagator yields the graviton exchange amplitude
in momentum space,
\bea
\frac{1}{2}\frac{1}{(2M_P)^{2}}\left( i\right) ^{2}\frac{-i}{q^{2}+i\epsilon }%
T^{\mu \nu } A_{\mu \nu \text{ }\rho \sigma } T^{\rho
\sigma }=\frac{1}{4M_P^{2}}\frac{-i}{q^{2}+i\epsilon }\left( 2T^{\rho \sigma
}T_{\rho \sigma }-TT\right) 
\eea
where $T=\eta ^{\rho \sigma }T_{\rho \sigma }$ {is the trace of the stress tensor.} This operator  corresponds
to the second term of the action, eq.(\ref{pot1}),
with the amplitude factor of $i$ (a combinatorial factor of $2$
will arise in a matrix element of this operator
in states such as $\langle m_1 m_2| ... |m_1 m_2\rangle$,
and reproduces the potential of eq.(\ref{newt})),

\subsection{Contact Terms from Single Graviton Exchange}

Here we evaluate the
operators
in the Feynman diagrams of Figure 1 arising from single graviton exchange
between $S_2$ and $S_3$. In classical background fields, $\phi_i $,
graviton exchange {between the pair} $\left\langle S_{2}S_{3}\right\rangle $ gives:
{\bea
-i\langle S_{2}S_{3}\rangle&=& -i(i^2) \int \int d^{4}y\;d^{4}x\;\frac{1}{(2M_{P})^{2}}F\left( x\right)  (-T^{\rho\sigma}(y))
\langle 0|\widehat{T} \text{ }\Pi ^{\mu \nu }h_{\mu \nu }(x)\text{ }h_{\rho \sigma }(y))\;|0\rangle
\nonumber\\
&=&\int \int d^{4}y\;d^{4}x\;\frac{1}{(2M_{P})^{2}}F\left(x\right)  \Pi ^{\mu \nu }A_{\mu \nu \rho \sigma
}D(x-y)T^{\rho \sigma}(y) 
\eea
}
where we have: 
\bea
\Pi ^{\mu \nu }A_{\mu \nu \rho \sigma }=2\partial ^{2}\eta
_{\rho \sigma }+4\partial _{\rho }\partial _{\sigma }.
\eea
Rearranging and integrating by parts:
\be
-i{\left\langle S_{2}S_{3}\right\rangle=\int \int d^{4}y\; d^{4}x\;\frac{F(x)}{2M_{P}^{2}}\left( \partial ^{2}D(x-y)T(y)-2D(x-y)\partial _{\rho }\partial _{\sigma }T^{\rho \sigma}\right)}
\ee
and we note that  $\partial _{\rho }\partial _{\sigma }T^{\rho \sigma}$
vanishes by the conservation of the 
stress {tensor.} The first term involving the trace, $T(y)$, is a contact term arising from
{$\partial^2 D(x-y)=\delta^4 (x-y)$.
Hence} the gravitational potential generates a contact term interaction in the 
effective action
of the form:
{\be
\int d^{4}x\; \frac{F\left( \phi _{i}( x) \right) }{2M_{P}^{2}}%
T( \phi _{i}(x)) 
\qquad\rightarrow \qquad
\int d^{4}x\;\frac{\xi \phi ^{2}}{2M_{P}^{2}}\left( -\partial ^{\mu
}\phi \partial _{\mu }\phi +4W(\phi )\right) 
\ee}
where we quote the general result and that of the simple model.

Furthermore, we have the exchange of a graviton involving
the pair $\left\langle S_{2}S_{2}\right\rangle $:
\bea
-i{\left\langle S_{2}S_{2}\right\rangle}&=&
-\frac{1}{2}i\left( i^{2}\right) \int \int d^{4}y\;d^{4}x\;
\frac{1}{%
(2M_{P})^2}F\left( x\right) F\left( y\right)
\langle 0|\widehat{T}\text{ }\Pi ^{\mu \nu }h_{\mu \nu }(x)\text{ }\Pi
^{\rho \sigma }h_{\rho \sigma }(y)|0\rangle   
\nonumber \\
&=&-\frac{1}{2}\int \int d^{4}y \; d^{4}x\; \frac{1}{4M_{P}^{2}}F\left( x\right)
\Pi ^{\mu \nu }A_{\mu \nu \rho
\sigma }\ \Pi ^{\rho \sigma }D(x-y)F\left( y \right). 
\eea
{Note the factor of $\frac{1}{2}$  coming }from the second order perturbative
expansion.
Here we have,
\bea
\Pi ^{\mu \nu }A_{\mu \nu \rho \sigma }\ \Pi ^{\rho
\sigma }=6\partial ^{2}\partial ^{2}
\eea
leading to the result:
\bea
-i{\left\langle S_{2}S_{2}\right\rangle}=-\int d^{4}x\;  \frac{3}{4M_{P}^{2}}\; F\left(
\phi _{i}\left( x\right) \right) \partial ^{2}F\left( \phi _{i}\left(
x\right) \right) .
\eea
This is the analogy of the $\frac{1}{2}A\partial ^{2}A$ term in the toy
model.

In summary the gravitational potential amongst $S_{2}$ and $S_{3}$ terms
mediated by a single graviton exchange diagram yields
contact terms that are an effective action, $S_{CT},$ and represents the
effect
of integrating out the $S_{2}$ term:
\bea
\label{CT}
S_{CT}= -\int d^{4}x\;  \frac{3}{%
4M_{P}^{2}} F\left( \phi _{i}\right) \partial ^{2}F\left( \phi
_{i}\right) +\int d^{4}x\;\frac{1 }{%
2M_{P}^{2}}F\left( \phi _{i}\right)T\left( \phi _{i}\right)
\label{SCT}
\eea
In the simple model case, we can rearrange the $F \partial ^{2}F $ term
to obtain,
\bea
S_{CT}=  \int d^{4}x\; 
\frac{3\xi ^{2}}{M_{P}^{2}}\; \phi ^{2}\partial \phi \partial \phi +
\int d^{4}x\;\frac{\xi \phi ^{2}}{2M_{P}^{2}}\left( -\partial ^{\sigma
}\phi \partial _{\sigma }\phi +4W(\phi )\right) 
\eea
Note the sign of the $F\partial ^{2}F$ is opposite (repulsive) to that of
the toy model, { a point that} we will clarify below.

\section{WEYL TRANSFORMATION}

In the previous section we directly evaluated the effective
action by calculating a single graviton exchange potential and
separating the contact terms, which must be interpreted as
parts of the effective action.  There is, however, another route,
which is to perform a Weyl transformation.

We can define: 
\bea
g_{\mu \nu }(x)=\Omega ^{-2}g_{\mu \nu }^{\prime }(x),\qquad g^{\mu
\nu }(x)=\Omega ^{2}g^{\mu \nu ^{\prime }}(x),\qquad \sqrt{-g}=\sqrt{%
-g^{\prime }}\Omega ^{-4}
\eea
and use:
\bea
R(\Omega ^{-2}g^{\prime })&=&\Omega ^{2}R(g)+6\Omega ^{3}D\partial
\Omega ^{-1}\nonumber\\
 L\left( g_{\mu \nu }(x),\phi _{i}(x)\right)&=&L\left( \Omega
^{-2}g_{\mu \nu }^{\prime }(x),\phi _{i}(x)\right)
\eea
With the choice $\Omega ^{2}=\left( 1+\frac{F\left( \phi_i \right) }{M_{P}^{2}}\right)$
we have:
\bea
S&\equiv &\int \sqrt{-g}\left(\frac{1}{2}M_{P}^{2}R\left( g_{\mu \nu }\right) +\frac{1}{2%
}F\left( \phi _{i}\right) R\left( g_{\mu \nu }\right) +L\left( \phi
_{i}\right) \right)\nonumber\\
&\rightarrow& \int \sqrt{-g^{\prime }}(\frac{1}{2}M_{P}^{2}R \left( g_{\mu \nu }^{\prime
}\right) +6\Omega D\partial \Omega ^{-1}+\Omega ^{-4}L\left( \Omega
^{-2}g_{\mu \nu }^{\prime }(x),\phi _{i}(x)\right) )
\eea
and we obtain:
\bea
S=\int \sqrt{-g^{\prime }}(\frac{1}{2}M_{P}^{2}
R\left( g_{\mu \nu }^{\prime }\right) -3M_{P}^{2}\partial_\mu \left( 1+\frac{F\left( \phi _{i}\right)}{%
M_{P}^{2}}\right) ^{+1/2}\partial^\mu 
\left( 1+\frac{F\left( \phi _{i}\right)}{M_{P}^{2}}\right)
^{-1/2} 
\nonumber \\
\qquad \qquad  +\left( 1+\frac{F}{M_{P}^{2}}\right) ^{-1}\frac{1}{2}%
g^{\mu \nu }\partial _{\mu }\phi \partial _{\nu }\phi -\left( 1+\frac{F\left( \phi _{i}\right)}{%
M_{P}^{2}}\right) ^{-2}W(\phi ,\chi ))
\eea
Keeping terms to $O({1\over M_P^2})$ and integrating by parts we have:
\bea
S=S_{1}+\int \left( L\left( \phi _{i}(x)\right) -\frac{3
F\left( \phi _{i}\right)\partial ^{2}F\left( \phi _{i}\right)}{4M_{P}^{2}}+\frac{F\left( \phi _{i}\right)T\left( \phi _{i}\right)}{2M_{P}^{2}}\right) 
\eea
The Weyl transformed action is identically consistent with the contact terms
of eq.(\ref{CT}) above,
to first order in $1/M_P^2$.

Hence, contact terms arise in gravity with non-minimal
couplings to scalar fields due to graviton
exchange. {Their form is equivalent to
a Weyl redefinition of the theory to one with a pure Einstein-Hilbert
action and reinforces their role as induced
components of the effective action.}  Hence working in any theory
with a non-minimal interaction $\sim F(\phi)R$ will lead
to these contact terms at order $1/M_P^2$. The contact terms can be
avoided in perturbation theory 
by going to the pure Einstein-Hilbert action with a Weyl tranformation.

The Weyl transformation is nonperturbative.
It is technically simpler  than the gravitational
potential
calculation, and it confirms the tricky normalization factors and phases in
the graviton
exchange calculation. As the Weyl transformation makes no reference to a gauge
choice, a calculation of the the contact terms in other gauges
should yield the
equivalent results. To check the invariance we turn now to a calculation in an alternative gauge which sheds further light on the origin of their structure.

\section{ANOTHER GAUGE}

Presently we will choose a gauge that will more clearly show what
is going on in the contact term equivalence with Weyl transformations. 
In particular, we obtained a negative sign for the analogy to the positive 
sign $A\partial ^{2}A$ of the toy  model, which becomes clear
in the present gauge choice.

We begin by defining traceless and trace fields for the weak field metric:
\bea
&&
s_{\mu \nu }=h_{\mu \nu }-\frac{1}{4}\eta _{\mu \nu }h
\qquad \qquad
t_{\mu \nu }=\frac{1}{4}\eta _{\mu \nu }h=\eta _{\mu
\nu }t
\eea
hence
$
h_{\mu \nu }=s_{\mu \nu }+\eta _{\mu \nu }t
$ and $ h=4t$.
The Fierz-Pauli action and non-minimal terms in these variables become,
\bea
S_{1} &=& \frac{1}{2}\int  \frac{3}{2}t\partial ^{2}t-\frac{1}{4}%
s^{\mu \nu }\partial ^{2}s_{\mu \nu }+\frac{1}{2}s^{\mu \nu }\partial _{\mu
}\partial ^{\rho }s_{\rho \sigma }-{3}t\partial _{\mu }\partial _{\nu }s^{\mu
\nu }
\nonumber \\
S_2 &=&
\int \frac{1}{M_{P}}F\left( \phi \right) \left( 3\partial ^{2}t\ -\partial
_{\lambda }\partial ^{\beta }s_{\beta }^{\lambda }\right) 
\eea
The coupling to gravity is:
\bea
 -\frac{h^{\mu \nu }}{2M_P}T_{\mu \nu }=-
\frac{s^{\mu \nu }}{2M_P}T_{\mu \nu }-\frac{t}{2M_P}T
\eea
Note that $t$ can be viewed as a small shift in the trace of the metric; $%
4\delta t=\delta h$,
and $\delta s=0$  and it therefore exclusively couples to the trace of the
matter field stress tensor.

Under a gauge transformation we have: 
\bea 
&&
\delta s_{\mu \nu }=\partial _{\mu }A_{\nu }+\partial
_{\nu }A_{\mu }-\frac{1}{2}\eta _{\mu \nu }\partial _{\rho
}A^{\rho }
\nonumber \\
&&
\delta t_{\mu \nu }=\frac{1}{2}\eta _{\mu \nu }\partial _{\rho
}A^{\rho }
\eea
Things simplify considerably if we can impose the gauge condition,
\bea
\partial ^{\mu }s_{\mu
\nu }=0.
\eea
Note that this is different from the condition $\partial ^{\mu
}h_{\mu \nu }=0$ owing
to the tracelessness of $s_{\mu \nu }\ .$ 
However, with $ \partial
^{\mu }s_{\mu \nu }=0$ we see that $s_{\mu \nu }$
exclusively 
contains the propagating modes of gravitational waves. 
For a gravitational wave propagating in the
$z-$direction in empty space the modes are $h_{xy}=s_{xy}$  and $
h_{xx}-h_{yy}=s_{xx}-s_{yy}$
and $t=0$.

Indeed, we can find a gauge transformation to fix
$
\partial ^{\mu }s_{\mu \nu }=0.$
Given any arbitrary configuration 
$s_{\mu \nu }^{0}$ and  $t_{\mu \nu }^{0}$  we can choose,
\bea
\partial ^{\mu }s_{\mu \nu }=\partial ^{\mu
}s_{\mu \nu }^{0}+\partial ^{2}A_{\nu }+\frac{1}{2}\partial
_{\nu }\left( \partial \cdot A\right) =0
\eea
and we find (see footnote 5):
\bea
A_{\nu }=-\frac{1}{\partial ^{2}}\partial ^{\mu }s_{\mu \nu
}^{0}+\frac{1}{3}\frac{\partial _{\nu }\partial ^{\rho }}{\partial
^{2}\partial ^{2}}\partial ^{\mu }s_{\mu \rho }^{0}
\eea
Verifying we see that,
\bea
\partial ^{\mu }s_{\mu \nu }=\partial ^{\mu
}s_{\mu \nu }^{0}+\partial ^{2}\left( -\frac{1}{\partial ^{2}}%
\partial ^{\mu }s_{\mu \nu }^{0}+\frac{1}{3}\frac{\partial
_{\nu }\partial ^{\rho }}{\partial ^{2}\partial ^{2}}\partial ^{\mu
}s_{\mu \rho }^{0}\right) -\frac{1}{3}\frac{\partial
_{\nu }\partial ^{\rho }}{\partial ^{2}}\partial ^{\mu }s_{\mu \rho }^{0}=0
\eea
Note that the gauge transformation also preserves
the traceless of $s_{\mu \nu }$ as,
\bea
\delta \eta ^{\mu \nu
}s_{\mu \nu }=2\partial^\nu  A_{\nu }-4\times \frac{1}{2}%
\partial _{\rho }A^{\rho }=0
\eea
Under this transformation we also redefine $t$:
\bea
t_{\mu \nu }=t_{\mu \nu }^{0}+\frac{1}{2}\eta _{\mu
\nu }\partial \cdot A=t_{\mu \nu }^{0}-\frac{1}{3}\eta _{\mu
\nu }\frac{1}{\partial ^{2}}\partial ^{\rho }\partial ^{\sigma }s_{\rho \sigma
}^{0}\qquad 
\eea
We remark that this gauge choice is one of  a single parameter, $w$, family
of gauge choices,
\bea
\partial_\mu h^{\mu \nu} = w \partial^\nu h
\eea
The De Donder gauge corresponds to $w=\frac{1}{2}$ while the present gauge choice,
$\partial_\mu s^{\mu \nu }=0$, corresponds to $w=\frac{1}{4}$.

In the $w=\frac{1}{4}$ gauge the Fierz Pauli action  simplifies to:
\bea
S_{1} &=&\frac{1}{2}\int \left( -\frac{3}{2}\partial t\partial t+\frac{1}{4}%
\partial s^{\mu \nu }\partial s_{\mu \nu }\right) 
\nonumber \\
&=&\frac{1}{2}\int \left( \frac{3}{2}t\partial ^{2}t-\frac{1}{8}s^{\alpha
\beta }\left( \eta _{\rho \alpha }\eta _{\sigma \beta }+\eta _{\sigma
\alpha }\eta _{\rho \beta }\right) \partial ^{2}s^{\rho \sigma }\right). 
\label{st}
\eea
The inverse of the kinetic term tensor is then,
\bea
&&
-\frac{1}{8}\left( \eta _{\mu \alpha }\eta _{\nu \beta }+\eta _{\mu \beta
}\eta _{\nu \alpha }\right) B^{\alpha \beta \text{ }\rho \sigma }=\frac{1}{2}%
\left( \delta _{\mu }^{\rho }\delta _{\nu }^{\sigma }+\delta _{\nu }^{\rho
}\delta _{\mu }^{\sigma }\right) 
\nonumber \\
&&
B^{\alpha \beta \text{ }\rho \sigma }=-2\left( \eta ^{\alpha \rho
}\eta ^{\beta \sigma }+\eta ^{\beta \rho }\eta ^{\alpha \sigma }\right) 
\eea
The propagator for $s_{\rho \sigma }$ is now,
\bea
\langle 0|\widehat{T}\;s_{\rho \sigma }\;s_{\alpha \beta
}\;|0\rangle  =\frac{-i}{q^{2}+i\epsilon }B_{\rho \sigma ,\alpha \beta }
\eea
The gauge invariance of amplitudes
is controlled by the conserved traceless tensors on the vertices.
Hence,  we must explicitly ensure that $s^{\mu \nu }$ couples
to conserved {\it and traceless} tensors only.
{Note that any conserved  field $s_{\mu \nu }$ can be made traceless, and
maintain conservation,
by applying the projection, }
\bea
s_{\mu \nu }\rightarrow s_{\mu \nu }-\frac{1}{3}\left( \eta _{\mu \nu }-
\frac{\partial _{\nu }\partial _{\mu }}{\partial ^{2}}\right) \eta ^{\rho \sigma }s_{\rho \sigma }.
\eea
Applying this to the energy momentum tensor the appropriate $s^{\mu\nu}$ coupling to a conserved and traceless stress tensor is given by,
\bea
&&
-\frac{s^{\mu \nu }}{2M_P}\widetilde{T}_{\mu \nu }-\frac{t}{2M_P}T\qquad 
\text{where} \qquad \widetilde{T}_{\mu \nu }=T_{\mu \nu }-\frac{1}{3}\left( \eta _{\mu
\nu }T-\frac{\partial _{\nu }\partial _{\mu }}{\partial ^{2}}T\right) .
\eea

We now repeat our calculation of the gravitational potential
in this gauge.  From the exchange of the $s^{\mu \nu }$ field with the
momentum space
projection operator on the vertices, 
\bea
\widetilde{T}_{\mu \nu }=T_{\mu \nu }-%
\frac{1}{3}\left( \eta _{\mu \nu }-\frac{q_{\mu }q_{\nu }}{q^{2}}\right) T
\eea
we have the amplitude,
\bea
\frac{1}{2}\ \left( \frac{1}{2M_P}\right) ^{2}{-i(i)^2\over q^2+i\epsilon}\widetilde{T}^{\rho
\sigma }\left( B_{\rho \sigma \alpha \beta }\right) \widetilde{T}^{\alpha
\beta }=\frac{-i}{q^{2}+i\epsilon }\left( \frac{1}{2M_P}\right) ^{2}\left( 
2T^{\rho \sigma }T_{\rho \sigma}-\frac{2}{3}TT \right) 
\eea
The exchange of the $t$ field which, {\it c.f.} eq(\ref{st}), has a noncanonical, and wrong sign for a scalar.
normalization and yields,
\bea
\label{kt}
\frac{-i}{q^{2}+i\epsilon }\left( i\right)^{2}
\frac{1}{3}\left( \frac{1}{2M_P}\right) ^{2}TT
\eea
and the sum of the $s$ and $t$ contributions is:  
\bea
\frac{-i}{q^{2}+i\epsilon }\left( \frac{1}{2M_P}\right)
^{2}\left( 2T^{\rho \sigma }T_{\rho \sigma }-TT\right) 
\eea
as obtained
previously in the De Donder gauge. The repulsive scalar
term, owing to the wrong sign kinetic term of $t$,
is absorbed into the full gauge invariant result,
and this reduces back to the De Donder gauge result 
which yields the Newtonian gravitational potential.
We therefore see explicitly that in two different gauges,
$w=\frac{1}{2}$ (De Donder) and $w=\frac{1}{4}$  ($\partial_\mu s^{\mu\nu}=0$
and $\eta_{\mu\nu}s^{\mu\nu}=0$)
the physical results are
equivalent. 
These are the two most interesting cases due to the simplifications detailed
above.

From eq.(\ref{st})
we see that the field $t$ is a ghost, however it is not produced
radiatively.
If we consider $F(\phi _{i})=0,$ then the equation of motion of $t$
is $\ 3\partial ^{2}t=\frac{2}{M_P}T\left( \phi _{i}\right) .$ However, we
can always write $T\left( \phi _{i}\right) $ as a divergence
of a current (the local scale or Weyl current,  $\partial ^{\mu }K_{\mu }=T$
\cite{FHR})
and therefore $\ 3\partial _{\mu }t=\frac{2}{M_P}K_{\mu }$ and $ t$  is
coupled in first
order to the Weyl current and becomes a ``tracker solution,''
$t=\int^x \frac{2}{3M_P}K_{\mu }dz^\mu $.  There is  no
 radiative wave, however there will be, e.g., cosmological solutions where $t$
 describes an expanding or shrinking universe.

However, the ghost field $t$ propagates off shell and will produce
a contact interaction. 
In this $w=\frac{1}{4}$
gauge the non-minimal term now depends only upon the trace field $t$:
\bea
\frac{1}{2}\int \sqrt{-g}FR=\int \frac{3}{2M_{P}}F \partial
^{2}t  
\eea
Consider the $ t$ part of the action,
\bea
S_{t}={1\over 2} \int -\frac{3}{2}\partial_\mu t\partial^\mu t+
\frac{3F}{2M_P}%
\partial ^{2}\frac{t}{M_P}-\frac{t}{2M_P}T.
\eea
We define a normalized field, $\chi$, by,
\bea
t=Z\chi \qquad \frac{3}{4}Z^{2}=\frac{1}{2}\qquad Z=\sqrt{\frac{2}{3}}
\eea
and,
\bea
S_{\chi}= \int \frac{1}{2}\chi \partial ^{2}\chi +\sqrt{\frac{3}{2}}%
\frac{F}{M_P}\partial ^{2}\frac{\chi }{M_P}-\frac{\chi }{2M_P}\sqrt{\frac{2}{3}}T
\eea
We see we essentially have the toy model action of {eq(\ref{toy})} but with the ghost sign for
the $\chi$ kinetic term. We can solve by shifting $\chi $:
\bea
\chi =\chi ^{\prime }-\sqrt{\frac{3}{2}}\frac{F}{M_P}+\frac{1}{\partial ^{2}}%
\sqrt{\frac{2}{3}}\frac{1}{2M_P}T
\eea
to obtain the contact terms,
\bea
S\rightarrow \int -\frac{1}{2}\partial \chi ^{\prime }\partial
\chi ^{\prime }+\frac{3}{4M_P}F\partial ^{2}F-\frac{F}{2M_P}T-\frac{1}{6M_P^{2}}T\frac{1}{\partial ^{2}}T.
\eea
where the large distance piece was computed above
and combines with the $s$ exchange to give the usual Newtonian
potential.  Restoring the original normalization
the effective action is therefore:
\bea
S=
\int -\frac{3}{2}\partial t\partial t+\frac{1}{4}\partial s^{\mu
\nu }\partial s_{\mu \nu }- \frac{3F\partial ^{2}F}{4M_{P}^{2}%
} +\frac{FT}{2M_{P}^{2}}+L(s,\phi _{i})-\frac{t}{2M_P}T\left( \phi
_{i}\right) -\frac{s^{\mu \nu }}{2M_P}\widetilde{T}_{\mu \nu }
\eea
The contact terms are the same as those found in eq(\ref{SCT}) in the De Donder gauge, demonstrating their gauge invariance as expected from the Weyl transformation structure.

\section{CONCLUSIONS}

We have provided some insight into the physical meaning and equivalence of actions related by a
Weyl transformation.  Our analysis confirms that  contact term effects are operant 
and that Weyl equivalent representations with non-minimal terms 
yield explicitly equivalent physics
to a pure minimal Einstein-Hilbert form.

The Weyl transformation to the minimal Einstein-Hilbert form
is, in a sense, inevitable. If one didn't know about the Weyl transformation
one would discover it in the induced contact
terms in the single graviton exchange potential involving non-minimal couplings. 
However the Weyl transformation is more powerful as it is fully non-perturbative.
Technically it provides a powerful check on the normalization and implementation
of the graviton propagators in various gauges, which can
otherwise be somewhat confusing.

The non-minimal form of the action is incomplete without including
the contact terms into the action.
The theory then becomes identical to the Weyl transformed form with a pure 
minimal Einstein-Hilbert
action. This implies that there are pitfalls in directly
interpreting the physics in the non-minimal form since the contact terms must
be included.  

The minimal Einstein-Hilbert action is special and
does not generate these contact terms.  In a sense, by going to the minimal Einstein-Hilbert
form we are diagonalizing the graviton derivative terms throught the action.
Our analysis required an Einstein-Hilbert term with a Planck mass and we expand
perturbatively in inverse
powers of $M^2_P$. A Weyl invariant theory, where $M_P=0$, is nonperturbative
and our analysis is then inapplicable. Indeed, there is
no conventional gravity in this limit since the graviton kinetic 
term does not then exist.  In this sense we view the formation of
the Planck mass by, e.g., inertial symmetry breaking, as a dynamical
phase transition, similar to a disorder-order phase transition in
a material medium \cite{FHR}. 

As an exploration of the gauge invariance of our result we have shown explicitly that, 
instead of the $w=1/2$, De Donder gauge, we can use
the $w=1/4$, $\partial_\mu s^{\mu\nu} =0$ gauge employing a traceless
$s_{\mu\nu} = h_{\mu\nu}-\frac{1}{4}\eta_{\mu\nu}h$ metric
together with a  separate trace field, $t$. 
A gauge transformation exists that takes arbitrary $s$ and $t$
to the $\partial_\mu s^{\mu\nu} =0$ gauge. Then we find that  $t$
exclusively controls the non-minimal term and the contact interations.
$t$ has a wrong sign (ghost) kinetic
term, however it is not produced as a propagating,
on shell gravitational wave. It nonetheless appears virtually
and, together with $s_{\mu\nu}$, produces the Newtonian potential
and the equivalent contact terms as obtained in De Donder gauge.

\vskip 0.5 in
\noindent
 {\bf Acknowledgements}
\vspace{0.1in}

We thank P.Ferreira and J. Noller for discussions.  Part of this work was done at 
Fermilab, operated by Fermi Research Alliance, 
LLC under Contract No. DE-AC02-07CH11359 with the United States 
Department of Energy. 

\vskip 0.15 in


\end{document}